# A preliminary study on a multi-resolution-level inverse planning algorithm for Gamma Knife radiosurgery

Zhen Tian, Xiaofeng Yang, Matt Giles, Tonghe Wang, Hao Gao, Elizabeth Butker, Tian Liu, Shannon Kahn

Department of Radiation Oncology, Emory University, Atlanta, GA 30022

Emails: Zhen.Tian@emory.edu

**Purpose**: Manual forward planning for GK radiosurgery is complicated and time-consuming, particularly for cases with large or irregularly shaped targets. Inverse planning eases GK planning by solving an optimization problem. However, due to the vast search space, most inverse planning algorithms have to decouple the planning process to isocenter preselection and sector duration optimization. This sequential scheme does not necessarily lead to optimal isocenter locations and hence optimal plans. In this study, we attempt to optimize the isocenter positions, beam shapes and durations simultaneously by proposing a multi-resolution-level (MRL) strategy to handle the large-scale GK optimization problem.

**Methods:** In our approach, several rounds of optimizations were performed with a progressively increased spatial resolution for isocenter candidate selection. The isocenters selected from last round and their neighbors on a finer resolution were used as new isocenter candidates for next round of optimization. After plan optimization, shot sequencing was performed to group the optimized sectors to deliverable shots supported by GK treatment units.

**Results:** We have tested our algorithm on 6 GK cases previously treated in our institution (2 meningioma cases, 3 cases with single metastasis and 1 case with 6 metastases). Compared with manual planning, achieving same coverage and similar selectivity, our algorithm improved the gradient index from 3.1±0.7 to 2.9±0.5 and reduced the maximum dose of brainstem from 8.0±4.3Gy to 5.6±3.8Gy. The beam-on time was also reduced by from 103.8±55.5 mins to 87.4±63.5 mins. Our method was also compared with the inverse planning algorithm provided in Leksell GammaPlan planning system, and outperformed it with better plan quality for all the 6 cases.

**Conclusions:** This preliminary study has demonstrated the effectiveness and feasibility of our MRL inverse planning approach for GK radiosurgery.

**Key words:** Gamma Knife radiosurgery, inverse planning, multi-resolution-level strategy

## 1. Introduction

Gamma Knife (GK) radiosurgery has emerged as an important and safe alternative to traditional neurosurgery to treat a variety of brain disorders [1-10]. Treatment planning is of utmost importance to achieve the desired treatment outcome. Currently, the most commonnly used planning approach for GK radiosurgery is manual forward planning, that is, planners manually place shots and adjust the isocenter location, collimator size for each of the eight sectors and beam duration for each shot [11-14]. With such many variables to adjust, manual planning is very challenging, which makes the resulting plan quality heavily depend on planners' skills, experiences and the amount of efforts invested in developing a plan.

Inverse planning may ease the GK planning process via solving an optimization problem. If multiple planning objectives are used in inverse planning, planners only need to adjust the priorities among them to meet the physicians preferred trade-off for each individual patient. This is similar to the inverse planning process for intensity modulated radiation therapy (IMRT) and volumetric modulated radiation therapy (VMAT). In addition, inverse planning also enables the planners to better exploit the full capabilities of the modern GK units. However, the vast search space involved in GK planning makes the optimization computationally expensive and usually exceed the capacity of the computing devices commonly used in clinics. For instance, roughly 5TB of memory is needed for inverse planning for a typical GK case[15]. To address this issue, a sequential planning strategy was employed by most of the existing GK inverse planning algorithms (including the algorithm provided in Leksell Gamma Plan (LGP), the commercial treatment planning system for the Leksell GK units)[11,13,14,16-19]. Specifically, based on the geometry of the targets, the isocenter locations are predetermined using a grassfire and sphere-packing algorithm or some other geometric methods[13,16,18]. Then the beam shapes and beam durations are optimized for these predetermined isocenter locations to achieve a good dose distribution. This sequential strategy substantially reduces the search space to ease the inverse planning. However, due to the dose interactions and contributions between neighboring isocenters, determining the isocenter locations is not exactly a geometric problem. The planning objectives and the physician preferred trade-off cannot be considered at the isocenter pre-selection stage due to the de-coupling of the inverse planning into two stages. Hence, the sequential planning strategy does not necessarily result in optimal isocenter positions and therefore optimal treatment plans.

In this paper, we attempt to optimize the isocenter positions, beam shapes and durations simultaneously. Ideally any coordinate within the target volume can be an isocenter candidate. A fine spatial resolution for the isocenter candidate grid is expected to offer more degrees of freedom and therefore a higher chance of finding an optimal solution. However, a fine spatial resolution would significantly increase the data size and make it exceed the capaciy of the computing devices. Inspired by the progressive resolution optimizer used in Eclipse treatment planning system for VMAT plan optimization[20], we proposed a multi-resolution-level (MRL) strategy to handle the large-scale GK optimization problem. Our hypothesis is that using a coarse resolution at the beginning should enable us to quickly explore different trade-offs that can be achieved for each particular patient to specify the physician's preference, and determine a rough distribution pattern of the optimal isocenters accordingly. Adding the neighbors of the optimized isocenters on a finer resolution into the next round of optimization should enable us to fine tune the isocenter locations to search for a better solution. We have performed a preliminary study to evuualuate the efficacy of our MRL inverse planning strategy for GK radiosurgery. The remaining part of this paper is organized as follows: Section 2 introduces our proposed method and implementation in details; Section 3 presents the planning results we have obtained for 6 GK patient cases; Sections 4 and 5 provide the conclusion and some discussions on our method.

## 2. Methods

The flowchart of our proposed MRL inverse planning framework is shown in Figure 1. It consists of two phases: an optimization phase and a shot sequencing phase. At the optimization phase, several rounds of optimization are performed with a progressively increased spatial resolution of the isocenter candidates. At each round, the beam-on time of each collimator and sector at each isocenter candidate is optimized. The shot sequencing phase is then to group those individual sectors with non-zero beam-on time to deliverable composite shots supported by the GK treatment unit[21]. The remaining of Section 2 will introduce each step of our method in details.

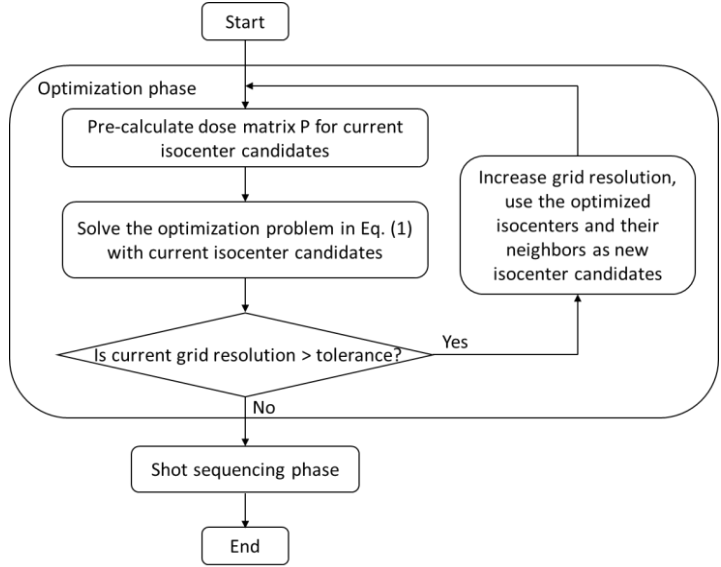

Figure 1. Flowchart of our proposed inverse planning framework

*2.1 The optimization phase*

*2.1.1 Calculating dose matrix P*

Before each round of optimization, it is necessary to calculate the dose rate at different voxels inside the patient skull contributed by each sector candidate (i.e. each of the eight sectors with each of the three available collimator sizes at each isocenter candidate). We employed our in-house developed GK second dose calculation engine for this dose calculation, which used tissue maximum ratio (TMR10) algorithm [22].

*2.1.2 Objective function*

It is reported in a recent study that linear programming tends to yield lower beam-on time than convex quadratic penalty approach and convex moment-based approach[23]. Hence we adopted the linear programming optimization model proposed by Nordström H originally for sector duration optimization on the preselected isocenters[14], and adapted it for our simultaneous optimization of isocenter location and sector duration optimization. The objective function used at each round of optimization with different sets of isocenter candidates is formulated as

$$\begin{aligned}\text{minimize}_t \ & \frac{\omega_{TH}}{N_T}\sum_{i\in I_T}\max(D_i-D_{TH},0) \ + \frac{\omega_{TL}}{N_T}\sum_{i\in I_T}\max(D_{TL}-D_i,0) \\ & +\frac{\omega_{IS}}{N_{IS}}\sum_{i\in I_{IS}}\max(D_i-D_{IS},0) \ +\frac{\omega_{OS}}{N_{OS}}\sum_{i\in I_{OS}}\max(D_i-D_{OS},0) \\ & +\frac{\omega_O}{N_O}\sum_{i\in I_O}\max(D_i-D_O,0) \\ & +\omega_{BOT}\sum_{n=1}^{N}\max_{m=1,2,\dots,8}\left(\sum_{c=1}^{3}t(c,m,n)\right)+\omega_S\|t\|_1,\end{aligned} \quad (1)$$

subject to $t \geq 0$.

Here, $t$ is the unknown beam-on time to be optimized for each sector candidate, that is, each sector ($m =1,2,\dots,8$) with each collimator size ($c =1, 2, 3$ corresponding to 4 mm, 8 mm and 16 mm collimators) at each isocenter candidate ($n =1,2,\dots, N$). $D_i = \sum_j p_{i,j}t_j$ denotes the total dose delivered to voxel $i$. The first six constraints in Eq.(1) are adopted from Nordström H's model. Specifically, we used target minimum dose and maximum dose to promote target coverage and dose homogeneity respectively (term 1 and 2 in Eq.(1)), the maximum dose received by the inner and outer shells of the target to control the selectivity and gradient index (term 3 and 4), the OAR maximum dose for OAR sparing (term 5), and the estimated total beam-on time (term 6) to penalize long treatment time. The seventh constraint is a sparsity constraint we added to enforce a penalty on the solutions with a large number of isocenters. Ideally, $l_0$-norm should be used on the vector $\sum_{m=1}^{8}\sum_{c=1}^{3}t(c,m,n)$ to count the number of the isocenters with non-zero beam-on time. However, this $l_0$-norm term would make the objective function non-convex. Our alternative way of approaching this problem is to replace the non-convex $l_0$-norm by an $l_1$-norm, which is the closest convex norm to the $l_0$-norm [24,25]. $\omega_{TH}, \omega_{TL}, \omega_{IS}, \omega_{OS}, \omega_O, \omega_{BOT}$ and $\omega_s$ denote the user-specified relative priorities among these constraints, which need to be adjusted to meet physician's trade-off preference. Note that only one target and OAR are considered in the objective function in order to simplify the expression of Eq.(1). Multiple targets and OARs can be handled simultaneously by our approach. Our objetive function can be expressed as a linear programming problem by introducing auxiliary variables (see Appendex A). We employed the dual simplex method to solve this linear programming problem.

*2.1.3 Multi-resolution-level (MRL) optimization strategy*

At the first round of our optimiztion, a coarse 3D grid was used for us to select all the grid points within the target volume as the isocenter candidates for the dose matrix calculation and the plan optimization modeled as Eq.(1). In our study, the priorities among the multiple planning objectives are tuned at this round to quickly explore the different trade-offs that can be achieved for each specific patient and then specify the trade-off preferred by the physician. At the next round of optimization, the isocenters with non-zero beam-on times were selected as new isocenter candidates. A finer grid resolution was used, and the grid points that were neighboring to these selected isocenters and within the target volume were added into the set of isocenter candidates. We then updated the dose matrix for these new isocenter candidates and rerun the optimization, with the set of priorities specified at the first round. This allowed the isocenters selected from last round to move around during the optimization to search for better solution. This process was repeated until reaching our resolution tolerance, which was set to be 0.5 mm in our study.

*2.2 The shot sequencing phase*

After determining the optimal isocenters and the beam-on times for the individual sectors at these isocenters, the sectors at a same isocenter need to be grouped into composite shots for delivery. In our work, we employed the shot sequencing algorithm developed by Nordström H *et al* [21], which tried to maximize the simultaneous delivery of radiation in order to minimize the total beam-on time. Specifically, for each of the eight sectors at an optimal isocenter, the collimator size that has the longest beam-on time is identified to consitute a composite shot. The beam-on time for this shot is set as the shortest non-zero beam-on time among these eight identified sectors. This procedure is repeated until all open sectors at that isocenter have been packaged. More details on shot sequencing can be found in [21]. In our study, the shots with beam-on time below 10 sec are removed to reflect the treatment unit limitations and reduce the effect of shutter dose .

*2.3 Postprocessing*

The LGP system is configured such that planners could only specify the relative weight of the shots, which is then used to calculate the beam-on time for each shot by the system. Hence, in order to put our plans back into the system, we need to convert our beam-on time to the relative weight for each shot. In our study, this relative weight of each shot was obtained by calculating the dose contribution of each shot to its own isocenter and normalizing these contributions such that the maximal value among all the shots equaled to one.

**3. Experimental Results**

*3.1 Materials*

Our MRL inverse planning algorithm was implemented in Matlab 2018b (The Mathworks, Inc.) on an Intel Core$^{TM}$ i9-7900x CPU with 64GB RAM. A resolution of 1 mm was used to determine the target voxels and the voxels of the inner and outer target shell for optimization. A coarser resolution of 2 mm was used to determine the OAR voxels. The spatial resolution used for the isocenter candidate at the highest level of our multi-resolution-level strategy was set to 0.5 mm.

**Table** 1. Case information.

| Case | Indication | Rx (Gy) | No. of targets | Target volume (cc) | Neaby OAR |
|---|---|---|---|---|---|
| 1 | Miningioma | 14 | 1 | 3.872 | brainstem |
| 2 | Miningioma | 14 | 1 | 2.612 | none |
| 3 | Metastase | 21 | 1 | 0.691 | none |
| 4 | Metastase | 15 | 1 | 10.929 | brainstem |
| 5 | Metastase | 21 | 1 | 2.056 | brainstem |
| 6 | Multiple Metastases | 15 | 6 | 0.701<br>0.792<br>0.393<br>1.674<br>0.519<br>0.867 | brainstem |

We have performed an IRB-approved restrospective study to validate our algorithm on six radiosurgery cases that have been previously treated in our institution. The information of these cases are specified in Table 1. The following guidelines, which were used in our clinic to guide manual forward planning for these GK cases, were adopted in our study to guide the inverse planning: (1) 100% of prescription dose $R_x$ must be received by at least 99% of the target volume, that is, $V_{100} \geq 99\%$; (2) Planner should try to maximize selectivity and minimize gradient index while achieving good target coverage; (3) The maximum dose to 0.1 cc of brainstem must not exceed 12 Gy. In our study, we compared the plans obtained by our algorithm with the original treatment plans that were created with manual forward planning and approved for clinical treatment. For comparison purpose, our clinical medical physcist also used the inverse planning module provided in LGP to create a plan for each case. Since both of our algorithm and the LGP inverse planning algorithm need users to adjust the priorities to specifiy their prefered trade-offs among the multiple objectives, for fair comparison purpose we tried to realize the similar trade-offs as reflected in the original manual plans during inverse planning. We would like to mention that because the dose matrix used in our algorithm was calculated by our in-house developed second dose engine, we input the plans obtained by our algorithm to LGP system to recalculate the dose for fair comparison.

The plan quality of the obtained plans are evaluated and compared in terms of maximum dose, target coverage (as defined in Eq.(2)), selectivity (Eq.(3)), gradient index (Eq.(4)), OAR maximum dose, and total beam-on time.

$$\text{Target coverage} = \frac{TV \cap PIV}{TV}. \tag{2}$$

$$\text{Selectivity} = \frac{TV \cap PIV}{PIV}, \tag{3}$$

$$\text{Gradient index} = \frac{PIV_{R_x/2}}{PIV}. \tag{4}$$

Here, TV and PIV represent the target volume and the planning isodose volume, respectively. $PIV_{R_x/2}$ represents the volume that receives at least half of the prescription dose.

*3.2 Results*

Our comparison results for cases 1-5 that have a single target are listed in Table 2. For cases 1-3, our MRL inverse planning algorithm achieved better plan qualities compared to manual forward planning, yielding same coverage, similar selectivity, slightly better gradient index and shorter beam-on time. In addition, much better sparing of brainstem was achieved by our algorithm for case 1. For cases 4-5, compared to manual forward planning, with the same coverage being obtained, our algorithm achieved slightly better selectivity, gradient index and sparing of brainstem, as well as much shorter beam-on time (i.e. 37.1 min versus 77.8 min for case 4 and 95.3 min versus 131.1 min for case 5). Although the plans obtained by the inverse planning algorithm in LGP for these five cases are clinically acceptable, they all have worse selectivity, gradient index, OAR sparing compared to our plans. For case 4, the beam-on time obtained by LGP inverse planning is two times as long as that obtained by our algorithm. For the other four cases, the beam-on times obtained by these two algorithms are comparable.

Table 2. Comparison of plan qualities for cases 1-5 that have a single target. The plans obtained by manual forward planning, the inverse planning algorithm in LGP and our inverse planning algorithm are compared. BOT represents the total beam-on time of the shots. For case 2 and 3, the brainstem dose was not applicable since brainstem was far away from target in these two cases and were not contoured.

| Case | Planning technique | Planning Isodose (%) | Max dose (Gy) | Coverage | Selectivity | Gradient index | Brainstem $D_{0.1cc}$ (Gy) | BOT (min) |
|---|---|---|---|---|---|---|---|---|
| 1 | Manual | 50 | 28.0 | 1.00 | 0.74 | 2.80 | 11.6 | 79.0 |
|   | LGP IP | 50 | 28.0 | 1.00 | 0.61 | 2.68 | 11.0 | 68.8 |
|   | Our IP | 50 | 28.0 | 1.00 | 0.73 | 2.57 | 4.2 | 64.8 |
| 2 | Manual | 50 | 28.0 | 1.00 | 0.84 | 2.51 | NA | 55.4 |
|   | LGP IP | 50 | 28.0 | 1.00 | 0.80 | 2.78 | NA | 48.1 |
|   | Our IP | 50 | 28.0 | 1.00 | 0.83 | 2.42 | NA | 47.9 |
| 3 | Manual | 50 | 41.9 | 1.00 | 0.68 | 3.18 | NA | 74.8 |
|   | LGP IP | 50 | 42.0 | 0.99 | 0.53 | 3.32 | NA | 69.0 |
|   | Our IP | 52 | 40.4 | 1.00 | 0.68 | 3.10 | NA | 68.9 |
| 4 | Manual | 50 | 30.0 | 1.00 | 0.70 | 2.70 | 3.3 | 77.8 |
|   | LGP IP | 50 | 30.0 | 1.00 | 0.60 | 2.95 | 4.8 | 72.8 |
|   | Our IP | 50 | 30.0 | 1.00 | 0.73 | 2.77 | 2.9 | 37.1 |
| 5 | Manual | 50 | 42 | 1.00 | 0.64 | 2.79 | 11.7 | 131.1 |
|   | LGP IP | 50 | 42 | 1.00 | 0.57 | 3.28 | 11.7 | 95.0 |
|   | Our IP | 50 | 42 | 1.00 | 0.67 | 2.68 | 11.2 | 95.3 |

The comparison result of the plans obtained for case 6 which has six targets are listed in Table 3. The isodose lines of the prescription dose and half of the prescription dose for each plan are shown in Figure 2. Because targets 2-5 in this case are very close to each other, the gradient index for each individual target is not applicable. Since all the six targets have same prescription dose, to better evaluate the plan quality we combined these targets into a single target and calculated the coverage, selectivity and gradient index. It is found that with same coverage and similar max dose obtained, our method has much better selectivity and gradient index, i.e. 0.65 selectivity and 3.77 gradient index compared to 0.57 and 4.51 obtained by manual planning and 0.47 and 5.11 obtained by LGP inverse planning. This is consistent with the much smaller area of the $R_x$ and $R_x/2$ isodose lines achieved by our method, as shown in Figure 2. Similar max dose, brainstem $D_{0.1cc}$ dose and total beam-on time were obtained by the three planning methods. We attribute the outperformance of our algorithm on this case to its ability of handling the multiple targets simultanesouly during optimization.

Table 3. Comparison of plan qualities for case 6 which has six targets. The plans obtained by manual forward planning, the inverse planning algorithm in LGP and our inverse planning algorithm are compared. The max dose and brainstem $D_{0.1cc}$ dose are reported for the composite plan that sums up the doses of the shots for all the targets. Because targets 2-5 are very close to each other, the gradient index for each individual target is not applicable. To better compare the plan quality, the six targets were combined to a single target to recalculate the coverage, selectivity and gradient index. BOT represents the total beam-on time of the shots.

| Target | Planning | Planning | Max dose | Coverage | Selectivity | Gradient | Brainstem | BOT |

| | technique | Isodose (%) | (Gy) | | | index | $D_{0.1cc}$ (Gy) | (min) |
|---|---|---|---|---|---|---|---|---|
| 1 | Manual | 50 | NA | 0.99 | 0.68 | 4.07 | NA | 25.7 |
| | LGP IP | 50 | NA | 1.00 | 0.69 | 3.79 | NA | 28.1 |
| | Our IP | 50 | NA | 1.00 | 0.75 | 3.11 | NA | 27.1 |
| 2 | Manual | 65 | NA | 1.00 | 0.48 | NA | NA | 10.7 |
| | LGP IP | 50 | NA | 1.00 | 0.45 | NA | NA | 26.0 |
| | Our IP | 50 | NA | 1.00 | 0.65 | NA | NA | 28.1 |
| 3 | Manual | 50 | NA | 1.00 | 0.51 | NA | NA | 30.4 |
| | LGP IP | 50 | NA | 1.00 | 0.45 | NA | NA | 30.1 |
| | Our IP | 51 | NA | 1.00 | 0.44 | NA | NA | 33.2 |
| 4 | Manual | 50 | NA | 1.00 | 0.46 | NA | NA | 80.6 |
| | LGP IP | 50 | NA | 1.00 | 0.32 | NA | NA | 67.1 |
| | Our IP | 51 | NA | 1.00 | 0.56 | NA | NA | 63.2 |
| 5 | Manual | 50 | NA | 1.00 | 0.59 | NA | NA | 20.3 |
| | LGP IP | 50 | NA | 1.00 | 0.60 | NA | NA | 19.9 |
| | Our IP | 50 | NA | 1.00 | 0.67 | NA | NA | 25.4 |
| 6 | Manual | 50 | NA | 1.00 | 0.54 | NA | NA | 37.0 |
| | LGP IP | 50 | NA | 1.00 | 0.34 | NA | NA | 27.3 |
| | Our IP | 51 | NA | 1.00 | 0.57 | NA | NA | 33.4 |
| Total plan (combined targets) | Manual | NA | 30.0 | 1.00 | 0.57 | 4.51 | 5.5 | 204.7 |
| | LGP IP | NA | 30.1 | 1.00 | 0.47 | 5.11 | 3.3 | 199.3 |
| | Our IP | NA | 30.0 | 1.00 | 0.65 | 3.77 | 4.0 | 210.5 |

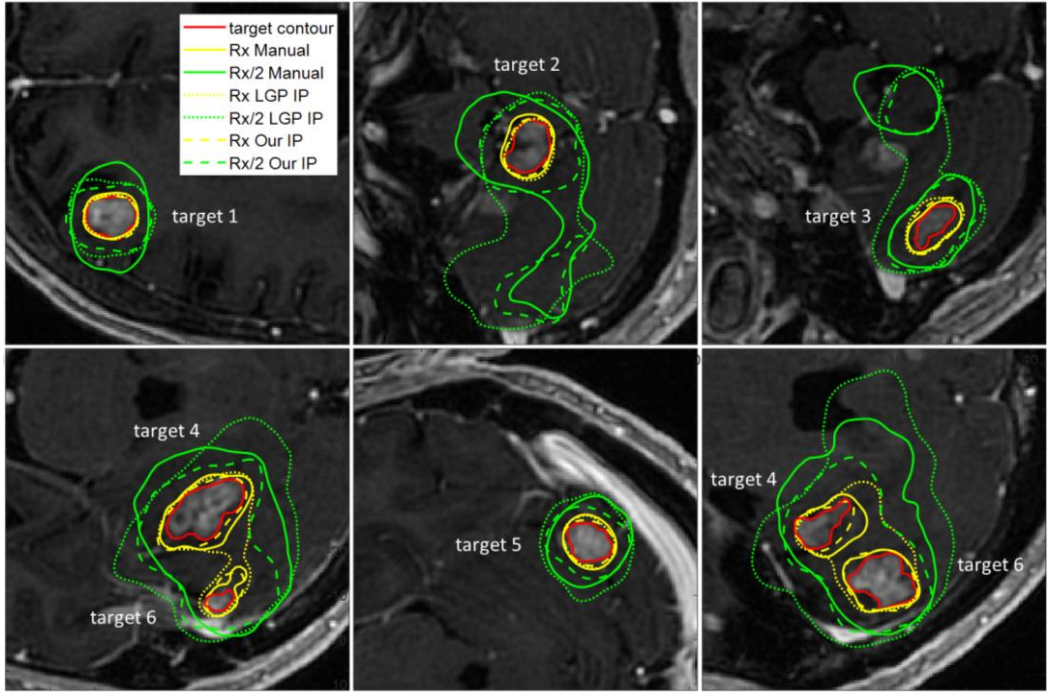

**Figure 2.** Isodose lines of the plans created for case 6 which has six targets. The target contours are shown in red. Isodose lines of the prescription dose $R_x$ and half of the prescription dose $R_x/2$ are shown in yellow and green, respectively. The isodose lines of the plans created by manual forward planning, LGP inverse planning and our inverse planning are shown in solid, short dashed and dashed curves, respectively.

To demonstrate the effectiveness of our proposed multi-level resolution strategy, Table 4 lists the amount of the isocenter candidates when directly using 0.5 mm grid resolution, as well as the amount of the isocenter candidates at each resolution level when using the multi-level strategy. The computation time for the corresponding optimization problems to converge are also listed in Table 4. Note that due to the much larger target size, case 4 adopted four levels of grid resolutions. All the other cases used three levels. The resolution at the highest level was 0.5 mm for all the six cases. It can be seen that this strategy substantially reduced the size of our optimization problem and hence the computation load. For instance, when directly using the 0.5 mm grid resolution, 5575 isocenter candidates needed to be considered in our inverse planning for case 3, which took 5706.7 seconds (~1.6 hours) for the optimization to converge. This long computation time made it clinically infeasible for planners to adjust the priorities among the objectives to explore the achievable trade-offs and specify their own preference. In contrast, with our multi-level resolution strategy, 74 isocenter candidates were considered at the first level for case 3 using a coarse resolution (i.e. 2 mm), which only took 3.6 seconds for the corresponding optimization at this level to get converged. In our experiments, we adjusted the priorities among the multiple objectives and specified our preference at this level for all the cases. These specified priorities were used for the other higher levels. When moving to the finer resolution at the next level, the isocenters selected in the optimization as well as their neighbors were used as the new set of isocenter candidates in the optimization. In case 3, we had 379 isocenter candidates at level 2 and 667 isocenter candidates at level 3, which took 49.8 seconds and 124.0 seconds to solve the corresponding optimization problems respectively. With 10 times of adjusting the priorities we had at the first level, the total computation time for case 3 was 209.8 seconds, much shorter than 5706.7 seconds when not using the multi-level strategy. User intervention was only needed for the fine tuning process. The computation time for the other five cases when not using the multi-level strategy is not available, because the optimization problem was either too huge to be solved on the computer used for this study or took longer time than our threshold (set to be 5 hours) to get converged and was hence terminated.

**Table 4.** Experimental results to demonstrate the effectiveness of the multi-level resolution strategy. This table lists the amount of isocenter candidates $N_{iso}$ and the corresponding computation time for the optimization to converge with/without using our multi-level resolution strategy. All the cases used three levels of grid resolutions, except case 4 which uses four levels of spatial resolutions. The resolution at the highest level was 0.5 mm for all the cases. The computation times for the corresponding optimization at each level to converge are also listed in this table.

|  | Case 1 | Case 2 | Case 3 | Case 4 | Case 5 | Case 6 |
|---|---|---|---|---|---|---|
| $N_{iso}$ when using 0.5 mm resolution without multi-level strategy | 31008 | 20911 | 5575 | 87500 | 16465 | 39878 |
| $N_{iso}$ for level 1 | 138 | 334 | 74 | 167 | 253 | 176 |
| $N_{iso}$ for level 2 | 603 | 705 | 379 | 566 | 629 | 967 |
| $N_{iso}$ for level 3 | 1221 | 915 | 667 | 954 | 850 | 2481 |
| $N_{iso}$ for level 4 | NA | NA | NA | 855 | NA | NA |
| Computation time without multi-level strategy (s) | NA | NA | 5706.7 | NA | NA | NA |
| Computation time for level 1 (s) | 236.6 | 158.4 | 3.6 | 22.1 | 150.0 | 36.3 |
| Computation time for level 2 (s) | 2013.7 | 491.3 | 49.8 | 163.5 | 804.8 | 375.5 |
| Computation time for level 3 (s) | 10517.9 | 923.4 | 124.0 | 467.4 | 1381.7 | 1977.4 |
| Computation time for level 4 (s) | NA | NA | NA | 430.6 | NA | NA |

## 4. Discussion and conclusions

In GK radiosurgery, the most commonnly used treatment planning approach is mannual forward planning, which is often complex, tedious and time consuming for large or irregularly shaped targets. Inverse planning may ease GK planning by solving an optimization problem. However, due to its vast search space, most of GK inverse planning algorithms adopt a sequential planning strategy to substantially reduce the problem size, which however does not necessarily result in optimal isocenter locations and hence optimal treatment plans. In this work, we have presented a multi-resolution-level strategy for GK inverse planning, attempting to optimize the shots' locations, sizes and beam-on time simultanesouly rather than in a sequential scheme. In our approach, several rounds of plan optimization were performed with a progressively increased spatial resolution of the isocenter candidates. At each round, the beam-on time of each collimator and sector at each isocenter candidate was optimized. The isocenters that had non-zero beam-on times as well as their neighbors on a finer grid resolution were selected to be the new isocenter candidates for the next round of optimization. After plan optimization, shot sequencing was then performed to group those individual sectors with non-zero beam-on time to deliverable composite shots supported by the GK treatment unit. Our inverse planning method has been tested on six patient cases, and achieved better plan quality than the original plans that were created by our GK physicians and physicists using manual forward planning and approved for clinical treatment. The results have also shown that our algorithm outperforms the sequential inverse planning module provided in LGP system, particularly for the cases with multiple targets that are close to each other.

A big difference of our algorithm from most of the other inverse planning algorithms are ours tries to optimize the shot location, shot shape and beam-on time simultaneously, instead of determining the isocenters and the beam shapes and durations sequentially. Although this can provide us a better chance to find an optimal solution, it substantially increases the problem size and computation load. The multi-level resolution strategy was proposed to reduce the size of the optimization problem to make it tractable. In our experiments on the six cases, it took 3.6~ 236.6 seconds to solve the optimization problem at level 1 with a coarse resolution, which made it feasible for planners to adjust the priorities among multiple optimization objetives to explore the different achievable trade-offs for each specific patient within a clinically acceptable time frame. We have found that the coarse resolution was good enough for fine tuning to explore the achievable trade-offs and specify the trade-off perfered by physicians, and the priorities specified at level 1 resulted in good plans at the higher levels for all the six cases we have tested. Using a finer resolution at the next round of optimization allows the isocenters selected from last round to move around during optimization to search for better solution. Although the computation time at the higher levels could be much longer than level 1 (e.g. 124.0 ~10517.9 seconds at the last level for the six cases), human invention is not needed at these levels, which enables planners to work on multiple cases in parallel. On the other hand, our algorithm is currently implemented in Matlab to preliminarily test the efficacy of our MRL-based simultaneous optimziation strategy for GK inverse planning. We will implement our algorithm on graphics processing units in the future, which is expected to solve the optimziation problem in real-time to further ease the planning process.

The disadvantage of the conventional sequential inverse planning strategy is that it decouples the plan optimization into two stages, which makes it impossible to consider the planning objetives and the physician's preferrence at the stage of isocenter preselection. Therefore, this sequential inverse planning doesn't necessarily generate optimal isocenter locations and hence optimal plans. Although our inverse planning approach consists of two phases, i.e. the plan optimization phase and the shot sequencing phase, our approach doesn't have this disadvantage. That's because although the deliverable shots are only available after the shot sequencing phase, the total beam-on time of these shots can be accurately calculated at the optimization phase and used as an penalty term in the plan optimization. Therefore, all the planning objectives and physician's preference are taken into account at the optimization phase to find an optimal plan.

This work is a preliminary study to demonstrate the effectivenss and feasibility of our inverse planning method for GK radiosurgery. Currently, the LGP system does not provide a data interface to import the shots' locations, shapes and relative weights (which are used to calculate shots' beam-on time in the system). We have to manually type these shot information into the LGP system, which greatly hinders the clinical applications of our method and also any other external GK planning algorithms. This issue can only be resolved by collaborating with Elekta in the future. Our future work is to employ artificial intelligence techniques to learn the physician's preferred trade-offs from previously treated patient cases and automate the fine tuning process of the priorities.

## Appendix

A linear programming problem is usually written in the following form:

$$\text{miniminze}_x\, f^T x,$$
$$\text{subject to } \begin{cases} A_{eq} x = b_{eq} \\ Ax \le b \\ lb \le x \le ub \end{cases}, \quad (A1)$$

where $x$ is the variable to be optimized. $f, b, b_{eq}, lb$ and $ub$ are vectors, and $A$ and $A_{eq}$ are matrices. To express the objective function in Eq.(1) to a linear programming problem, auxiliary variables need to be introduced. For instance, two nonnegative variables $v^+$ and $v^-$ are introduced which satisfy $D_i - D_0 = v^+ - v^-$, so that the one-side penalty term $\max(D_i - D_0, 0)$ can be rewritten as $v^+ \ge D_i - D_0$ and $\max(D_0 - D_i, 0)$ as $v^- \ge D_0 - D_i$. To handle the penalty term enforced on the beam-on time, i.e. $\max_{m=1,2,\ldots,8}(\sum_{c=1}^{3} t(c,m,n))$, in linear programming, two non-negative variables $\tau$ and $d$ are introduced such that $\tau(n) = d(m,n) + \sum_{c=1}^{3} t(c,m,n)$. With these auxiliary variables, our objective function can be expressed as:

$$\text{miniminze}_x\, f^T x,$$
$$\text{subject to } \begin{cases} A_{eq} x = b_{eq} \\ x \ge 0 \end{cases}.$$

where,

$$x = (V_{TH}^+, V_{TH}^-, V_{TL}^+, V_{TL}^-, V_{IS}^+, V_{IS}^-, V_{OS}^+, V_{OS}^-, V_O^+, V_O^-, d, \tau, t),$$
$$f^T = (W_{TH}, 0, 0, W_{TL}, W_{IS}, 0, W_{OS}, 0, W_O, 0, 0, W_{BOT}, W_S),$$
$$b_{eq}^T = (D_{TH}, D_{TL}, D_{IS}, D_{OS}, D_O, 0),$$

$$A_{eq} = \begin{pmatrix} -I & I & 0 & 0 & 0 & 0 & 0 & 0 & 0 & 0 & 0 & 0 & P_T \\ 0 & 0 & -I & I & 0 & 0 & 0 & 0 & 0 & 0 & 0 & 0 & P_T \\ 0 & 0 & 0 & 0 & -I & I & 0 & 0 & 0 & 0 & 0 & 0 & P_{IS} \\ 0 & 0 & 0 & 0 & 0 & 0 & -I & I & 0 & 0 & 0 & 0 & P_{OS} \\ 0 & 0 & 0 & 0 & 0 & 0 & 0 & 0 & -I & I & 0 & 0 & P_O \\ 0 & 0 & 0 & 0 & 0 & 0 & 0 & 0 & 0 & 0 & I & -H & G \end{pmatrix}, \quad (A2)$$

$$G = I_{8N} \otimes I_{1\times 3} = \begin{pmatrix} 1 & 1 & 1 & 0 & 0 & 0 & \cdots & 0 & 0 & 0 \\ 0 & 0 & 0 & 1 & 1 & 1 & \cdots & 0 & 0 & 0 \\ \vdots & \vdots & \vdots & \vdots & \vdots & \vdots & \ddots & \vdots & \vdots & \vdots \\ 0 & 0 & 0 & 0 & 0 & 0 & \cdots & 1 & 1 & 1 \end{pmatrix} \in \mathcal{R}^{8N \times 24N},$$

$$H = I_N \otimes I_{8\times 1} = \begin{pmatrix} I_{8\times 1} & 0 & \cdots & 0 \\ 0 & I_{8\times 1} & \cdots & 0 \\ \vdots & \vdots & \ddots & \vdots \\ 0 & 0 & \cdots & I_{8\times 1} \end{pmatrix} \in \mathcal{R}^{8N \times N}.$$

$\otimes$ denotes the Kronecker product.

The variable $x$ is composed of 13 variables. Except the beam-on time of each sector $t$ to be optimized, the other 12 variables are all auxiliary variables. $V_{TH}^+, V_{TH}^-, V_{TL}^+, V_{TL}^-$ are the non-negative auxiliary variables used to handle the penalty terms imposed on the target voxels, and their lengths are all equal to the amount of the target voxels used in optimization. $V_{IS}^\pm, V_{OS}^\pm, V_O^\pm$ are the non-negative auxiliary variables, with the length equal to the amount of the inner shell voxels, the amount of the outer shell voxels and the amount of the OAR voxels that are used in optimization, respectively. $f^T$ is also composed of 13 vectors, which are the specified priorities corresponding to each component of $x$. Specifically, $W_{TH}$ is a row vector whose length is equal to the amount of the target voxels used in optimization, with the value of each element being $\omega_{TH}$. $W_{TL}$, $W_{IS}$, $W_{OS}$, $W_O$ are constructed similarly. $W_{BOT}$ is a row vector with a length of $N$ and the value of each element is $\omega_{BOT}$. $W_S$ has $24N$ elements whose values are all $\omega_S$.

**Reference**

1.  Barnett GH, Linskey ME, Adler JR, et al. Stereotactic radiosurgery—an organized neurosurgery-sanctioned definition. *Journal of neurosurgery.* 2007;106(1):1-5.
2.  Flickinger JC, Kondziolka D, Lunsford LD, et al. A multi-institutional experience with stereotactic radiosurgery for solitary brain metastasis. *International Journal of Radiation Oncology* Biology* Physics.* 1994;28(4):797-802.
3.  Kondziolka D, Lunsford LD, Coffey RJ, Flickinger JC. Stereotactic radiosurgery of meningiomas. *Journal of neurosurgery.* 1991;74(4):552-559.
4.  Petrovich Z, Yu C, Giannotta SL, O'day S, Apuzzo ML. Survival and pattern of failure in brain metastasis treated with stereotactic gamma knife radiosurgery. *Journal of neurosurgery.* 2002;97(Supplement 5):499-506.
5.  Schneider BF, Eberhard DA, Steiner LE. Histopathology of arteriovenous malformations after gamma knife radiosurgery. *Journal of neurosurgery.* 1997;87(3):352-357.
6.  Karlsson B, Lindquist C, Steiner L. Prediction of obliteration after gamma knife surgery for cerebral arteriovenous malformations. *Neurosurgery.* 1997;40(3):425-431.
7.  Kondziolka D, Zorro O, Lobato-Polo J, et al. Gamma Knife stereotactic radiosurgery for idiopathic trigeminal neuralgia. *Journal of neurosurgery.* 2010;112(4):758-765.
8.  Kondziolka D, Lunsford LD, Flickinger JC, et al. Stereotactic radiosurgery for trigeminal neuralgia: a multiinstitutional study using the gamma unit. *Journal of neurosurgery.* 1996;84(6):940-945.
9.  Lunsford LD, Niranjan A, Flickinger JC, Maitz A, Kondziolka D. Radiosurgery of vestibular schwannomas: summary of experience in 829 cases. *Journal of neurosurgery.* 2005;102(Special_Supplement):195-199.
10. Wu A. Physics and dosimetry of the gamma knife. *Neurosurgery Clinics of North America.* 1992;3(1):35-50.
11. Wu QJ, Chankong V, Jitprapaikulsarn S, et al. Real‐time inverse planning for Gamma Knife radiosurgery. *Medical physics.* 2003;30(11):2988-2995.
12. Levivier M, Carrillo RE, Charrier R, Martin A, Thiran J-P. A real-time optimal inverse planning for Gamma Knife radiosurgery by convex optimization: description of the system and first dosimetry data. *Journal of neurosurgery.* 2018;129(Suppl1):111-117.
13. Ghobadi K, Ghaffari HR, Aleman DM, Jaffray DA, Ruschin M. Automated treatment planning for a dedicated multi‐source intracranial radiosurgery treatment unit using projected gradient and grassfire algorithms. *Medical physics.* 2012;39(6Part1):3134-3141.
14. Sjölund J, Riad S, Hennix M, Nordström H. A linear programming approach to inverse planning in Gamma Knife radiosurgery. *Medical physics.* 2019.
15. Ghaffari HR, Aleman DM, Jaffray DA, Ruschin M. A tractable mixed-integer model to design stereotactic radiosurgery treatments. *Technical Report MIEOR‐TR2012‐12.* 2012.
16. Wu QJ, Bourland JD. Morphology‐guided radiosurgery treatment planning and optimization for multiple isocenters. *Medical Physics.* 1999;26(10):2151-2160.
17. Schlesinger DJ, Sayer FT, Yen C-P, Sheehan JP. Leksell GammaPlan version 10.0 preview: performance of the new inverse treatment planning algorithm applied to Gamma Knife surgery for pituitary adenoma. *Journal of neurosurgery.* 2010;113(Special_Supplement):144-148.

18. Wagner TH, Yi T, Meeks SL, et al. A geometrically based method for automated radiosurgery planning. *International Journal of Radiation Oncology* Biology* Physics.* 2000;48(5):1599-1611.

19. Oskoorouchi MR, Ghaffari HR, Terlaky T, Aleman DM. An interior point constraint generation algorithm for semi-infinite optimization with health-care application. *Operations research.* 2011;59(5):1184-1197.

20. Kan MW, Leung LH, Yu PK. The performance of the progressive resolution optimizer (PRO) for RapidArc planning in targets with low‐density media. *Journal of applied clinical medical physics.* 2013;14(6):205-221.

21. Nordström H, Johansson J. Sequencing sector fields. In.: Google Patents; 2015.

22. Elekta Instrument A. *A new TMR dose algorithm in Leksell GammaPlan.* Technical Report;2011.

23. Cevik M, Ghomi PS, Aleman D, et al. Modeling and comparison of alternative approaches for sector duration optimization in a dedicated radiosurgery system. *Physics in Medicine & Biology.* 2018;63(15):155009.

24. Donoho DL. Compressed sensing. *IEEE Transactions on information theory.* 2006;52(4):1289-1306.

25. Emmanuel C, Romberg J, Tao T. Robust uncertainty principles: Exact signal reconstruction from highly incomplete frequency information. 2004.